\documentclass[prl,aps,showpacs,twocolumn,superscriptaddress]{revtex4}
\usepackage{latexsym}
\usepackage{graphicx}

\newcommand{\be}{\begin{equation}}
\newcommand{\ee}{\end{equation}}
\newcommand{\ba}{\begin{eqnarray}}
\newcommand{\ea}{\end{eqnarray}}
\newcommand{\baa}{\begin{eqnarray*}}
\newcommand{\eaa}{\end{eqnarray*}}

\begin{document}

\title{Collinear to Anti-collinear Quantum Phase Transition by Vacancies}
\author{Bao Xu}
\affiliation{Beijing National Laboratory for Condensed Matter
Physics, Institute of Physics, Chinese Academy of Sciences, Beijing
100080, China}
\author{Chen Fang}
\affiliation{Department of Physics, Purdue University, West
Lafayette, Indiana 47907, USA}
\author{W.M. Liu}
\affiliation{Beijing National Laboratory for Condensed Matter
Physics, Institute of Physics, Chinese Academy of Sciences, Beijing
100080, China}
\author{Jiangping Hu}
 \affiliation{Department of Physics,
Purdue University, West Lafayette, Indiana 47907, USA}
\affiliation{Beijing National Laboratory for Condensed Matter
Physics, Institute of Physics, Chinese Academy of Sciences, Beijing
100080, China}

 \begin{abstract}
    We study static vacancies in the collinear magnetic phase of  a
    frustrated Heisenberg $J_1$-$J_2$ model. It is found  that vacancies
    can rapidly suppress the collinear antiferromagnetic state (CAFM)
    and generate a new magnetic phase, an anti-collinear magnetic phase
    (A-CAFM), due to magnetic frustration. We investigate the quantum
    phase transition between these two states by studying a variety of
    vacancy superlattices. We argue that the anti-collinear magnetic
    phase can exist in iron-based superconductors  in the absence of any
    preceding structural transitions and an observation of this novel
    phase will unambiguously resolve the relation between the magnetic
    and structural transitions in these materials.
 \end{abstract}

 \pacs{74.25.Ha,74.40.Kb,74.70.Xa}

 \maketitle

  There are several reasons for studying static vacancy problems on frustrated magnetic systems.
    First of all,
  there has been convincing experimental evidence which supports the magnetism in
  iron-based high temperature superconductors ($Fe$-HTSC) can be understood by an effective frustrated magnetic
  model ($J_1$-$J_2$-$J_z$ model) \cite{Fang2008d,Si2008,Xu2008a,Yildirim2008} which simultaneously captures
  the collinear antiferromagnetic state and  the tetragonal to
  orthohombic structural transition  observed in neutron-scattering experiments \cite{Cruz2008}.
  The new superconductors are very flexible in substituting  $Fe$ by other transition metal atoms,
  such as $Mn$, $Zn$, $Co$ and $Ni$.
  The static-vacancy problem in the $J_1$-$J_2$-$J_z$ model is, then, an important low energy effective
  model for non-magnetic $Zn$-doped $Fe$-HTSC \cite{Cheng2010}.
  Moreover, the recently discovered  122 iron-chalcogenide,
  $(K, Cs) Fe_{2-x}Se_2$ \cite{Guo2010,Fang2010,Liu2011},
  carries intrinsic iron vacancies,
  which can even form superlattice vacancy structures
  \cite{Bao2011a,Bacsa2011,Pomjakushin2011b,WangZ2011,Zavalij2011,ZhangAM2011}.
  Thus, the solution of the static-vacancy problem can be directly tested experimentally and contributes to a
  fundamental understanding regarding the role of magnetism in superconductivity as well as the
  coupling between lattice and magnetism.
    Second,
  with  various frustrated magnetic materials being discovered
  in the past decade, many novel physics and new states of matter  have been proposed.  However, experimentally,
  it has  often been difficult to identify  features associated to novel physics, for example,
  spin liquid state \cite{Shimizu2003, Lee2005}. Static vacancies can either enhance or decrease the degree of
  frustration and can behave rather differently in different state of matters.
  Therefore, static vacancies can  contribute to a new understanding of frustrated magnetic
  physics and provide unique features that can be probed experimentally.
    Finally,
  even in a standard quantum Heisenberg antiferromagnetic model, it has been shown that quantum
  fluctuations
  can also be dramatically modified around   static vacancies \cite{Bulut1989,Chen2010}.
  Studying  static vacancies in frustrated
  quantum magnetic systems can also provide a deeper understanding
  of the interplay between quantum fluctuations and geometric frustration.

    In this Letter, we study the static vacancy problem in the
    $J_1$-$J_2$ antiferromagntic Heisenberg model.   We employ a linear
    spin-wave (LSW) theory \cite{Bulut1989} to understand properties of
    a single static vacancy and static vacancy superlattices.  We show,
    depending on the frustrated coupling, quantum fluctuations can be
    either reduced or enhanced on neighbors of an isolated vacancy. More
    importantly, by calculating the exact ground-state properties of a
    variety of static vacancy lattices, we predict that sufficient
    static vacancies can cause a quantum phase transition between the
    collinear magnetic phase and an anti-collinear magnetic phase before
    a spin glassy phase without a spatial long-range magnetic order
    is formed.

\begin{figure}
     \includegraphics[ scale=0.35 ]{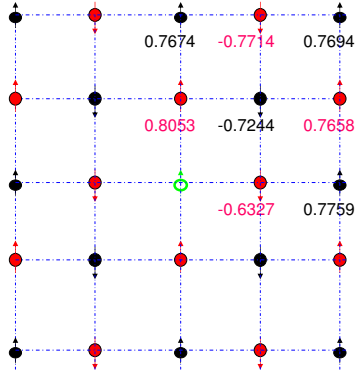}
     \caption{ (color online) The sketch of the CAFM state in the $J_1-J_2$ model
     where the two sublattices A and B are colored by
      black and red respectively and a single vacancy site is at the center.
      The numbers labels the magnetic order parameter $<S_z(i)>$ at each site
     in the CAFM state in the presence of a single vacancy for the parameters
     $J_1=J_2$ and $S=1$ (without the vacancy, the
     magnetic order parameter is $|<S_z(i)>|=0.7817$).  }\label{fig1}
    \end{figure}

    Without vacancies, the $J_1$-$J_2$ model is given by
     \begin{eqnarray}
     H_{0}=J_{1}\sum_{<ij>_{NN}}  \hat S_{i}\cdot
     \hat S_{j}+J_{2}\sum_{<ij>_{NNN}}\hat S_{i}\cdot
     \hat S_{j}, \label{ham}
     \end{eqnarray}
     where $<ij>_{NN}$ and $<ij>_{NNN}$ denote bonds formed by two nearest neighbor sites
     and two next nearest neighbor sites respectively.
     For a classical $J_1$-$J_2$ model with $J_1<2J_2$, the ground state can be viewed
     as two decoupled antiferromagnetically (AFM) ordered states
     on the A and B sublattices as shown in Fig.\ref{fig1}.
     Including quantum fluctuations, the relative angle between the two antiferromagnetic
     orders on the A and B sublattices is locked and the quantum model
     has a CAFM ground state with an ordering wave vector at $Q=(0,\pi)$ or $Q'=(\pi,0)$.
     The CAFM state is driven by the frustrated coupling $J_1$.
     The energy of quantum fluctuations can be
    calculated using the standard LSW theory.
    Without losing generality,  we take the AFM order in the A sublattice as $ S^z_A\neq 0$ and
    the AFM order in the B sublattice
    rotates  by $\theta$ around y-axis relative to the one in the A sublattice. Namely,
     \begin{eqnarray}
     \left( \begin{array}{c}S^{B}_{x}\\ S^{B}_{y}\\ S^{B}_{z}\end{array}\right)
     =\left(\begin{array}{ccc}{\rm cos}\theta&&{\rm sin}\theta\\&1&\\-{\rm sin}\theta&&{\rm cos}\theta
     \end{array}\right)
     \left( \begin{array}{c}S^{A}_{x}\\ S^{A}_{y}\\
     S^{A}_{z}\end{array}\right),
     \end{eqnarray}
    In  the   LSW  approximation, Eq.(\ref{ham}) reduces to
     \begin{eqnarray}
    H_{0}
    &=&\sum_{k}\omega_{k} b^{\dagger}_{k} b_{k}+\frac{\nu_{k}}{2}( b_{-k} b_{k}+{\rm
    h.c.})-4NJ_{2}S^{2},\label{hank}
    \end{eqnarray}
    where,
    $\omega_{k}=S[J_{1}{\rm cos}\theta({\rm cos}k_{x}-{\rm cos}k_{y})+J_{1}({\rm cos}k_{x}+{\rm
    cos}k_{y})+4J_{2}]$,
    $\nu_{k}=S[J_{1}{\rm cos}\theta({\rm cos}k_{x}-{\rm cos}k_{y})-J_{1}({\rm cos}k_{x}+{\rm
    cos}k_{y})-4J_{2}{\rm cos}k_{x}{\rm cos}k_{y}]$,
    and $b$ ($ b^+$) are magnon annihilation (creation) operators.
    Using the  Bogliubov transformation,
    $\hat b_{k}={\rm cosh}\psi_k  \alpha_{k}+{\rm
    sinh}\psi_k  \alpha^{\dagger}_{-k},$
    Eq.(\ref{hank}) can  be diagonalized as
    \begin{eqnarray}
    H_{0}=\sum_{k}\tilde{\omega}_{k}\alpha^{\dagger}_{k}\alpha_{k}+N\epsilon_{0}-4NJ_2S(S+1),
    \end{eqnarray}
    where,
    $
     \tilde{\omega}_{k}=\sqrt{\omega^{2}_{k}-\nu^{2}_{k}},
    $

    $
    {\rm cosh}^{2}\psi=\frac{1}{2}\Big[ \frac{\omega_{k}}{\sqrt{\omega^{2}_{k}-\nu^{2}_{k}}}+1\Big]$,
    ${\rm sinh}^{2}\psi=\frac{1}{2}\Big[ \frac{\omega_{k}}{\sqrt{\omega^{2}_{k}-\nu^{2}_{k}}}-1\Big]$,
    and
    \begin{eqnarray}
    \epsilon_{0}=\frac{1}{N}\sum_{k}\frac{1}{2}\sqrt{\omega^2_k-\nu^2_k}=c-a_0\frac{J^2_1S}{J_2}{
    \rm cos}^2\theta,
    \end{eqnarray} where $c$ is independent of $\theta$ and $a_0\simeq 0.033$.
    $\epsilon_0 $ describes the well-known ``order by disorder mechanism''
    and has a minimum at $\theta =0,\pi$, favoring a CAFM order
    \cite{Henley1989,Chandra1990,Fang2008d}.

    For a single vacancy at the origin of the A-sublattice, the total
    Hamiltonian can be written
    as
  $H=H_0-V$ with
    $ V= J_1\sum_{i_{NN}}\hat S_0\cdot\hat  S^B_i+J_2\sum_{j_{NNN}}\hat
    S_0\cdot\hat  S^A_{j}. $
    In the LSW approximation, the Hamiltonian becomes
    \begin{widetext}
    \begin{eqnarray}
   H
    =\sum_{k}\tilde{\omega}_{k}\alpha^{\dagger}_{k}\alpha_{k}
 -\frac{1}{N}\sum_{k}C_{k}(\alpha_{k}+\alpha^{\dagger}_{k})+(N-1)\epsilon_{0}
 -\frac{1}{N^{2}}\sum_{\vec{k},\vec{q}}
 \left[\widetilde{A}_{k,q}\alpha^{\dagger}_{k+q}\alpha_{k}
 +\frac{B_{k,q}}{2}(\alpha^{\dagger}_{-k+q}\alpha^{\dagger}_{k}+\alpha_{-k+q}\alpha_{k})\right].
 \end{eqnarray}
 \end{widetext}
 where
 $\widetilde{A}_{k,q}=A_{k,q}{\rm cosh}(\psi_{k+q}+\psi_{k})+\nu_{k}{\rm sinh}(\psi_{k+q}+\psi_{k})$,
$ B_{k,q}=A_{k,q}{\rm sinh}(\psi_{k+q}+\psi_{k})+\nu_{k}{\rm cosh}(\psi_{k+q}+\psi_{k})$, and
 $C_{k}=\mu_{k}({\rm cosh}\psi_{k}+{\rm sinh}\psi_{k})$ with $
  \mu_{k}=-2SJ_1\sqrt{S/2}{\rm sin}\theta({\rm cos}k_x-{\rm cos}k_y)$.
 Up to the first  order of $J_1^2/J_2$, the total ground state energy in the presence of a single vacancy
 is $
 E_{0}=-4(N-1)J_2S(S+1)+(N-1)\epsilon_{0}+\epsilon_{v}$, where
 \begin{eqnarray}
 \epsilon_{v}=-\frac{1}{N}\sum_{k}\frac{C^{2}_{k}}{\sqrt{\omega^{2}_{k}-\nu^{2}_{k}}}=-a_1 \frac{J_1^2S^2}{J_2}
 {\rm sin}^2\theta,
 \end{eqnarray}
 where $a_1\simeq 0.36$.
 $\epsilon_{v}$ has a minimum at $\theta=\pm \pi/2$ and does not
 favor a CAFM state.  The physics behind the energy $\epsilon_{v}$ can be argued as follows. In the CAFM state,
 creating a vacancy at  one sublattice is similar to applying an external magnetic field
 along magnetic ordered direction on the four neighbor sites of the vacancy in the other
 sublattice. Since  the spins of the four neighbor sites are AFM,
 the presence of such a field would favor the AFM order
 in the four neighbor sites to be perpendicular
 to the external magnetic field direction.  $\epsilon_0$ and
 $\epsilon_v$ have different dependence on the spin $S$. The
 competition between these two energies can lead to a new phase
 transition.

    Considering the model with  a small density of vacancies, $\rho$, in
    the first order approximation and up to a constant, we can
    approximate the energy density of the model as a function of
    $\theta$ to be
    \begin{eqnarray}
     \epsilon(\theta,\rho)=(1-\rho)\epsilon_{0}(\theta)+\rho \epsilon_v(\theta).
    \end{eqnarray}
    The energy density favors the CAFM state($\theta=0,\pi$) if
    $\rho<\rho_c$ and an A-CAFM state ($\theta=\pm \pi/2$) if
    $\rho>\rho_c$ where the critical vacancy density is given by
    \begin{eqnarray} \rho_c=\frac{a_0}{a_0+a_1 S}.\end{eqnarray}
    Pluging in the values of $a_0$ and $a_1$, we obtain $\rho_c= 0.086$
    for $S=1$ and $\rho_c= 0.158 $ for $S=1/2$. These critical values are
    well below the percolation threshold which destroys the long range
    AFM order.

    We can also solve the single vacancy problem exactly (within the LSW approximation).
    Defining  the standard Green functions:
    \begin{eqnarray}
    & & G_{j,j'}(t)=-i<T[b_j(t)b_{j'}^+(0)]> \nonumber \\
    & & F_{j,j'}(t)=-i<T[b^+_{j}(t)b_{j'}^+(0)]>,
    \end{eqnarray}
    and their Fourier transformation
    $G(F)_{j,j'}(t)=\frac{1}{N^{2}}\sum_{k}\sum_{q}
    e^{i\vec{q}\cdot\vec{r}_{j}}e^{-i\vec{k}\cdot(\vec{r}_{j'}-\vec{r}_{j})+i\omega t}G(F)_{k+q,k}$,
    we can derive the following dynamic equations for the Green
    functions in the presence of a single vacancy at the origin of the lattice,
    \begin{widetext}
    \begin{eqnarray}\label{green}
    G_{k+q,k}&=&G^{0}_{k}\delta_{q,0}+\frac{1}{N}\sum_{p}\left[
    A_{k+q,p}(G^{0}_{k+q}G_{p,k}+F^{0}_{k+q}F_{p,k})
    +B_{k+q,p}(G^{0}_{k+q}F_{p,k}+F^{0}_{k+q}G_{p,k})\right],\nonumber\\
    F_{k+q,k}&=&F^{0}_{k}\delta_{q,0}+\frac{1}{N}\sum_{p}\left[
    A_{k+q,p}(\bar{G}^{0}_{k+q}F_{p,k}+F^{0}_{k+q}G_{p,k})
    +B_{k+q,p}(\bar{G}^{0}_{k+q}G_{p,k}+F^{0}_{k+q}F_{p,k})\right],
    \end{eqnarray}
    \end{widetext}
    where $G(F)^{0}$ are given by
    \begin{eqnarray}
    \left(\begin{array}{c}G^{0}_{k}\\F^{0}_{k}\end{array}\right)
    =\frac{1}{\omega^{2}-\tilde{\omega}^{2}_{k}}
    \left(\begin{array}{c}\omega+\omega_{k}\\-\nu_{k}\end{array}\right),
    \end{eqnarray}
    and $ A_{k+q,p}=J_{1}S\Big(-2{\rm cos}\theta[{\rm
    cos}(q_{x}+k_{x}-p_{x})-{\rm cos}(q_{y}+k_{y}-p_{y})]
    +{\rm cos}\theta[{\rm cos}(q_{x}+k_{x})-{\rm cos}(q_{y}+k_{y})+{\rm cos}p_{x}-{\rm
    cos}p_{y}]
    +[{\rm cos}(q_{x}+k_{x})+{\rm cos}(q_{y}+k_{y})+{\rm cos}p_{x}+{\rm
    cos}p_{y}]\Big)
    +4J_{2}S[1+{\rm cos}(q_{x}+k_{x}-p_{x}){\rm cos}(q_{y}+k_{y}-p_{y})
    ]$,
    $B_{k+q,p}=J_{1}S\Big( {\rm cos}\theta[{\rm cos}(q_{x}+k_{x})-{\rm cos}(q_{y}+k_{y})
    +{\rm cos}p_{x}-{\rm cos}p_{y}]
     -[{\rm cos}(q_{x}+k_{x})+{\rm cos}(q_{y}+k_{y})+{\rm cos}p_{x}+{\rm cos}p_{y}]\Big)
    -4J_{2}S[{\rm cos}p_{x}{\rm cos}p_{y} + {\rm cos}(q_{x}+k_{x}){\rm
    cos}(q_{y}+k_{y})]$.
    The Dyson-type equations in Eqs.(\ref{green}) can be solved
    numerically for any given $\theta$ and $J_1/J_2$ values. We focus on
    the magnetic order moments and the total energy on  sublattices
    surrounding the vacancy located at the origin $(0,0)$.

    First, in Fig.\ref{fig1}, we report the magnetic order parameter
    $<S_z(i)>$ at each site
    in the CAFM state ($\theta=0$) for the parameters  $J_1=J_2$ and $S=1$ (without the vacancy, the
    uniform  magnetic order  is $|<S_z(i)>|=0.7817$). In
    Fig.\ref{fig1}, we plot the magnetic moments at the sites $(0,1)$,
    $(1,0)$ and $(1,1)$ as a function of $J_1/J_2$ in the CAFM state.
    There are two important results: (1) the effects of the vacancy on
    its nearest neighbor (NN) sites are different along the two
    directions in the CAFM state. The zero-point deviations are
    suppressed (enhanced) at the NN sites along the ferromagnentic (AFM)
    directions if $J_1$ is AFM and the results reverse if $J_1$ is
    negative (ferromagnetic); (2) the effect of the vacancy on its next
    nearest neighbor (NNN) does not break $C_4$ rotation symmetry even
    in the CAFM state.  The zero-point deviation at these sites is
    suppressed for small $|J_1|$ values. This result is not surprising
    since it is known to be the case for $J_1=0$. However, the deviation
    goes from depression to enhancement as $|J_1|$ increases further.
    This crossover reflects the frustration increases the transverse
    fluctuations due to the anti-collinear tendency between the two
    magnetic moments of the sublattices around the vacancy.

\begin{figure}[]
    \includegraphics[ scale=0.7 ]{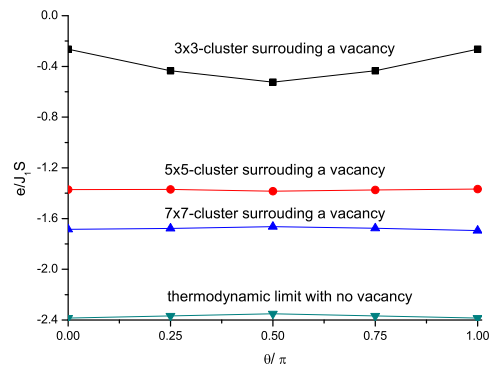}
    \caption{ (color online) The $\theta$-dependence of {\it energy per spin}
    in the presence of single vacancy for three different
    size of clusters: $3\times3$, $5\times5$ and $7\times7$. The
    parameters are chosen as
    $J_{2}=J_{1}=1$ and $S=1$.}\label{cluster3x3}
    \end{figure}

\begin{figure}[]
    \includegraphics[width=8cm ]{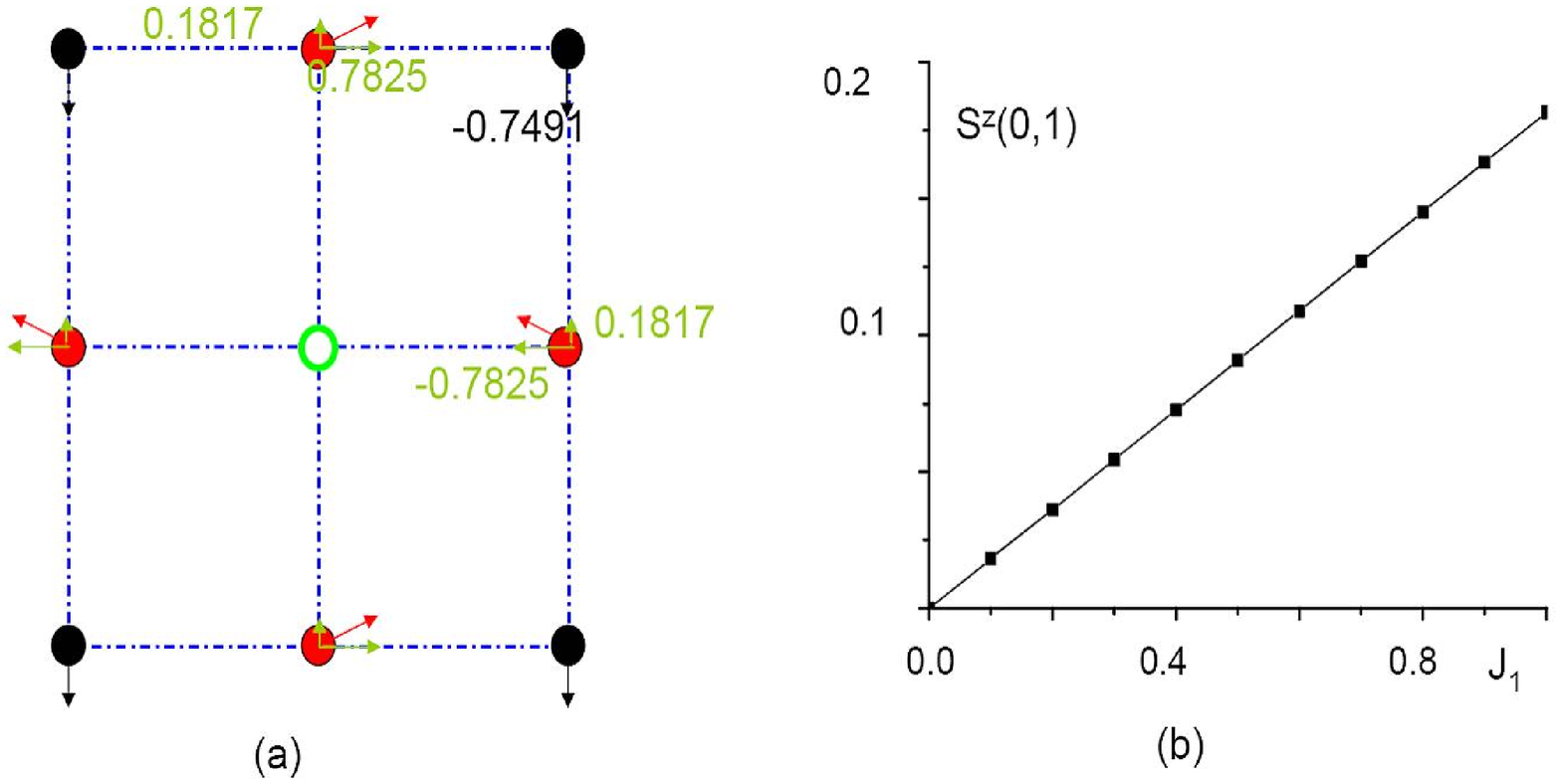}
    \caption{ (color online) (a) The magnetic moment configuration around the vacancy in
    $3\times 3$ cluster with $J_1=J_2=1$ and $S=1$. (b)
    The $J_{1}$-dependence of $S^{z}_{(1,0)}$ and $S^{z}_{(0,1)}$ for
    $S=1$ and $J_{2}=1$.  }\label{fig3}
    \end{figure}

    Second, we calculate the total energy of the model on clusters
    centered at the static vacancy as a function of $\theta$.  In
    Fig.\ref{cluster3x3}, we plot the energy on three different clusters
    surrounding the vacancy with sizes, $3\times 3$, and $5\times 5$ and
    $7 \times 7$ and parameters $S=1, J_1=J_2$. It is clear that the
    energy minimum for a $3\times 3$ cluster is $\theta=\pi/2$.
    Moreover, the static magnetization at the $8$ sites in the $3\times 3$
    lattice is around $0.9 \mu_b$ for $\theta =\pi/2$ which is larger
    than the case with no vacancies $0.798\mu_B$ at the CAFM phase. This
    result confirms that the vacancy clearly favors an A-CAFM ordering
    between two sublattices.  In Fig.\ref{fig3}(a), we plot the
    configuration of magnetic moment surrounding the vacancy in the
    A-CAFM state $\theta=\pi/2$ with $J_1=J_2=1$ and $S=1$. The value of
    the magnetic moment along the z direction for the nearest neighbour
    site of the vacancy  linearly increases as a function of $J_1$ as
    shown in Fig.\ref{fig3}(b).

\begin{figure}[]
    \includegraphics[ scale=0.6 ]{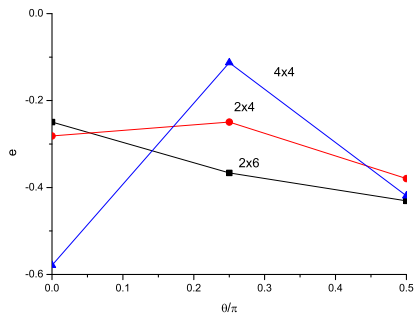}
    \caption{ (color online) The energy of three different superlattices as a function of
    $\theta$ for $J_1=J_2$. For the $4\times 4$ lattice,
    the energy minimum becomes $\theta=0$.}\label{vacsuper}
    \end{figure}

    The above study of a single vacancy suggests that the A-CAFM
    configuration is favored if only the energy on the small sublattice
    surrounding the vacancy is considered. In order to confirm that the
    existence of the global phase transition    in the presence of
    vacancies, we take a super unit cell in the square lattice with
    $N\times M$ sites and creates one vacancy in the unit. Thus, if we
    repeat this unit to create a superlattice, we obtain a system in
    which the percentage of vacancy concentration is given by
    $\frac{1}{N\times M}$. In this superlattice system, for a given
    wavevector $k$, the  Eq.(\ref{green}) can be reduced to equations that
    only couple $2N\times M$ Green functions given by $G_{k+Q_{n,m},k}$
    and $F_{k+Q_{n,m},k}$, where $Q_{n,m}=(2\pi n/N,2\pi m/M)$ and $n(m)
    =0,...,N(M)-1.$ In Fig.\ref{vacsuper}, we show the energy of three
    different supperlattices  as a function of $\theta$ for $J_1=J_2$
    and $S=1$. For both superlattices with $2\times 4$ and $2\times 6$
    unit cells which are corresponding to  $12.5\%$ and $8.4\%$ vacancy
    concentration respectively, the anti-collinear state is favored.
    However, the CAFM state is favored in a superlattice with $4\times
    4$ unit cell corresponding to $6.3\%$ vacancy concentration. This
    result justifies our previous rough estimation of the critical
    vacancy density.

    Our above results have important implications in iron-based
    superconductors. All of our above calculations demonstrate an
    existence of quantum phase transition from a  CAFM state to an
    A-CAFM state at a certain critical vacancy concentration $\rho_c$.
    While the CAFM state breaks $C^4$ rotation symmetry, the A-CAFM
    state does not break $C^4$ rotation symmetry. In  iron-pnictides,
    there is always a tetragonal-to-orthorhombic structural transition
    which occurs at the temperature above or equal to CAFM transition
    temperature.  This structural transition breaks $C^4$ to $C^2$ and
    is naturally explained as a consequence of magnetic fluctuations
    associated with the CAFM state \cite{Fang2008d,Xu2008a}. If the
    A-CAFM state exists and the structural transition is  magnetically
    driven, our results predict that the lattice distortion can be
    absent in the A-CAFM phase.

    It is also worth to discuss that the vacancy orderings have been
    observed in $(A_{1-y}Fe_{2-x}Se_2)$ iron-chalcogenides, where the
    vacancy patterns are corresponding to a natural reduction of the
    magnetic frustration so that the magnetic transition temperature is
    strongly enhanced \cite{Fang2011d}.  The vacancy superlattices used
    in our calculation do not reduce the magnetic frustration.
    Therefore, our results do not directly apply to the observed vacancy
    patterns, such as the 245 pattern in $K_2Fe_4Se_5$ \cite{Bao2011a}.
    However, for the materials with very diluted  vacancy concentration,
    we expect that our result should be valid as well.

    In summary, we study static vacancies in the collinear magnetic
    phase of  a frustrated Heisenberg $J_1$-$J_2$ model and identify a
    quantum phase transition between  collinear antiferromagnetic state
    (CAFM) and  an anti-collinear antiferromagnetic phase (A-CAFM). Our
    results can help to resolve the relation between magnetic and
    structural transitions in iron-based superconductors.

    {\it Acknowledge} JPH thanks S. Kivelson for initiating the main
    idea in this paper and for useful guide and discussion.
    This work was supported by NSFC under grants Nos. 10874235, 10934010, 60978019,
    the NKBRSFC under grants Nos. 2009CB930701, 2010CB922904, and 2011CB921502.

\end{document}